\documentclass[dvips,a4paper,10pt]{article}
\usepackage{graphicx}
\usepackage{multicol}
\newcounter{mysection}
 \newcommand{\mysection}[1]
    {\vspace{12pt}\begin{center}{{\large\Roman{mysection}}.\hspace{2pt}
    {\textbf{\large  #1}}\\} \end{center}\indent
     \addtocounter{mysection}{1}}
\newcounter{mysubsection}
 \newcommand{\mysubsection}[1]
    {\vspace{12pt}\begin{center}{{\large\Alph{mysubsection}}.
    {\textbf{#1}}\\} \end{center}\indent
     \addtocounter{mysubsection}{1}}
\newenvironment{changemargin}[2]{%
\begin{list}{}{%
\setlength{\leftmargin}{#1}%
\setlength{\rightmargin}{#2}%
\setlength{\listparindent}{\parindent}%
\setlength{\itemindent}{\parindent}%
\setlength{\parsep}{\parskip}%
}%
\item[]}{\end{list}}
\newcommand{\vT}{\textit{v}(T)\hspace{1pt}}
\newcommand{\aT}{$\alpha$(T)}

\begin{document}
  \author{\small Vasilii Gusakov \footnote{gusakov@ifttp.bas-net.by}
  \\[-5pt]\textit{\small Institute of Solid State \& Semiconductor Physics,%
  P. Brovki str. 17, 220072 Minsk, Belarus}}
  \title{\large Abnormal temperature dependence of elastic properties of cuprates superconductors}
   \date{}
\maketitle
  \begin{changemargin}{1cm}{1cm}
 {\small\vspace{-20pt}
 In cuprates superconductors (as polycrystal and single crystal samples) unusual
 hysteretic temperature behavior of elastic properties is frequently observed.
 In this  work from the uniform point of view detailed analysis this abnormal
 temperature behavior of elastic properties is given. It is shown that the hysteretic
 temperature dependencies of elastic properties of YBa$_2$Cu$_3$O$_{7-\delta}$ crystal are strongly
 anisotropic and are connected to hysteretic behavior of the $C_{3333}$ modulus  while the
 $C_{1111}$ and $C_{2222}$ moduli  have no such dependence. The analysis of the elastic
 constant tensor on the basis of the microscopic model has allowed to draw a conclusion, that
 the hysteretic behavior of the $C_{3333}$ modulus of YBa$_2$Cu$_3$O$_{7-\delta}$ crystal is
 caused by temperature dependent renormalization of constants of interaction of apex oxygen
 atoms with copper atoms and are connected apparently to formation  of charge ordering.}
\end{changemargin}
\vspace{10pt}
  \begin{multicols}{2}
      \mysection{Introduction}
 Already the first investigations of high temperature
superconductors  \cite{1,2,3,4} have revealed  a variety of essential features in the
temperature dependencies of velocity (\vT) and attenuation (\aT) of ultrasound. So for
the most well investigated isostructural series of HTSC compounds
(RE)Ba$_2$Cu$_3$O$_{7-\delta}$ (RE - rare earth atom)  in the temperatures region 190
- 230 K step changes of \vT have been observed \cite{5,6}, indicating on lattice
instabilities and/or phase transitions at temperatures previous to the superconducting
transition. Surprising there was that at heating and cooling  of a sample the
temperature dependencies of velocity (attenuation) did not coincide. A distinct
thermal hysteresis was observed between 60 and 230 K \cite{7,8,9}. Further
investigations have shown that the hysteretic behavior of \vT  is observed for all
isostructural series (RE)Ba$_2$Cu$_3$O$_{7-\delta}$ \cite{10,11}, and also for the Bi
\cite{12,13,14,15}, Pb \cite{16}, Tl \cite{17} HTSC compounds. Also appeared that for
polycrystal samples the value of  the hysteresis depends on structure of polycrystal
samples. So hysteretic behavior of \vT  was observed in textured polycrystal samples
\cite{18} and polycrystal samples with relatively large crystal grains ($\sim$50
$\mu$m) \cite{19}. In fine grinding, dense polycrystal samples \cite{20} and also
polycrystal samples obtained by  MPMG (melt-powder-melt-growth) method \cite{21} such
features were not observed. The results marked above specified a possible connection
of anomalous temperature dependencies \vT with physical processes proceeding on
crystal grain boundaries and, hence, for single crystals hysteretic behavior should
not be observed. However, further investigations have revealed the hysteretic behavior
of \vT in (RE)Ba$_2$Cu$_3$O$_{7-\delta}$ single crystals \cite{22,23} . So in
\cite{22} for ultrasound with frequency 10 MHz  the hysteretic behavior of the
C$_{3333}$ and C$_{1212}$ elastic constants was observed , while hysteretic behavior
of C$_{1111}$, C$_{2222}$, C$_{1313}$ and C$_{1212}$ was not observed.
 When studying of elastic properties of single crystals  YBa$_2$Cu$_3$O$_{7-\delta}$ at ~ 100 ÊHz  hysteretic behavior of both elastic
 and torsion modulus were observed \cite{23}. This  fact disagrees with
 results of \cite{22}, where  the hysteretic behavior of the C$_{1313}$
 and C$_{1212}$ modules  was not observed (module of torsion is determined by these
  elastic constants).
Observation in single crystals  of hysteretic behavior of \vT puts under doubt  the
models offered before   for an explanation of these dependencies (see, for example,
\cite{19}). Really, it is generally believed that in cuprate  superconductors
anomalous temperature behavior of ultrasonic wave velocity  is result from
redistribution of oxygen atoms  in the Cu-O chains  or/and phase transitions in
twinning area of crystals \cite{24,25}. In this connection it is necessary to note,
that the hysteretic behavior of velocity of ultrasound also was observed in
(RE)Ba$_2$Cu$_4$O$_{8}$ polycrystals \cite{26,27}. This material is stable down to 850
$^o$C and crystals are untwined. Besides while investigating of a linear expansion
coefficient of  YBa$_2$Cu$_3$O$_{7-\delta}$  single crystals  in \cite{28} was shown
that in chains disordering processes of oxygen atoms proceed at temperatures above 300
$^o$C. Let's pay also attention that the hysteretic behavior of \vT recently was
observed by us and in Ba$_x$K$_{1-x}$BiO$_{3}$  compound \cite{29}. For
Ba$_x$K$_{1-x}$BiO$_{3}$ and (RE)Ba$_2$Cu$_4$O$_{8}$ compounds the hysteretic behavior
of ultrasound can not be closely related to redistribution of oxygen atoms or
twinning. Earlier we offered a phenomenological model describing hysteretic
temperature behavior of elastic constants of HTSC single crystals \cite{30}. However
the microscopic mechanism resulting in so anomalous temperature behavior of velocity
of ultrasound and essential distinction  in elastic properties of polycrystal and
single crystal samples of HTSC compounds remains obscure till now.
 The solution of this task is important for the following reasons.
Recently for an explanation of the phase diagram of cuprate superconductors (remaining
mysterious after 13 years of research \cite{31}) the local micro-strain of Cu-O bonds
and charge ordering (formation of striped phases) \cite{31} were used. The
micro-strain of Cu-O bonds should result in appreciable changes of elastic properties
of crystal. Besides, hysteretic behavior of wave velocity of ultrasound was discovered
recently in $ \rm MgB_2$ compound \cite{32}. The huge temperature hysteresis of
velocity of a sound and internal friction was observed also in $\rm
La_{0.8}Sr_{0.2}MnO_3$ single crystals \cite{33}.

 In the given work is shown that the
hysteretic behavior of velocity of ultrasound, as in   single crystal, and in
polycrystal samples of coprate superconductors  may be explained from the uniform
point of view. In a final part the possible microscopic mechanism of hysteretic
behavior of ultrasound is discussed.
 \mysection {Anomalous temperature dependencies of elastic properties of single crystals}
 To develop  a detailed microscopic
model of hysteretic behavior of ultrasonic velocity and to analyze temperature
dependencies of ultrasonic wave velocity in polycrystal materials we must know the
components of elastic tensor which have anomalous temperature dependence. However to
the present time there are only two experimental works devoted to detailed analysis of
temperature dependencies of elastic properties of YBa$_2$Cu$_3$O$_{7-\delta}$  single
crystals but as I have marked in introduction, from \cite{22} and \cite{23} directly
it is impossible to make an unambiguous inferences about the temperature dependencies
of elastic modules. First of all  I  show that the experimental results of \cite{22}
are consistent with results \cite{23} quantitatively in the assumption, that anomalous
temperature dependence the elastic modulus C$_{3333}$ has only.

 Let's calculate resonant frequencies of a sample at deformation of
a torsion and compression in view of inertia of cross movement of elements of a
sample. We will consider the case when the laboratory system of coordinates
 is  oriented arbitrarily in reference to the crystallographic system of coordinates
 and the calculations are feasible by a variation method.
   \mysubsection { Orientations    $\vec u \bot \vec c,\vec k||\vec c$;
       $\vec k \bot \vec c,\vec u||\vec c$}
 For the given experimentally investigated in [23] orientations  of laboratory and
 crystallographic systems of coordinates
 ($\vec{u}$ is the displacement vector of an ultrasonic wave; $\vec{c}$ is the unit
 vector of the crystallographic direction [001]; $\vec{k}$  is the wave vector) in a sample
 the deformation of a torsion and a bend is created.
For evaluation of  natural frequencies of oscillations of a sample as generalized
coordinates it is convenient to pick the tensor  $\Omega _{ij}  =
{\raise0.7ex\hbox{${\partial \varphi _i }$} \!\mathord{\left/
 {\vphantom {{\partial \varphi _i } {\partial x_j }}}\right.\kern-\nulldelimiterspace}
\!\lower0.7ex\hbox{${\partial x_j }$}} $  (see, for example, \cite{35}), which is
closely related to the deformation tensor ($u_{ij}$ ) $\Omega _{ij}  = \delta _{ikl}
\frac{{\partial u_{lj} }}{{\partial x_k }}$ \cite{36}. If the tensor $\Omega_{ij}$ is
known, that the deformation may be restored to within moving a sample as the whole
\cite{37}. When measuring natural frequencies  of a sample  value $\Omega_{ij}$ is
usually small and constant. For small values of  $\Omega_{ij}$ we can expand a vector
and tensor of deformation into a Taylor series of $\Omega_{ij}$  and take into account
linear terms only
\begin{eqnarray}
 {u_i  = B_{ijk} (x_1 ,x_2 ,x_3 )\Omega _{jk} ;}\nonumber \\
   {u_{ij}  = A_{ijkl} (x_1 ,x_2 ,x_3 )\Omega _{kl}} ,
 \end{eqnarray}
 Then the free energy of the deformed sample may be written as
follows \begin{eqnarray}\label{2}
  F = \frac{1}{2}\int\limits_V {C_{ijkl}u_{ij}u_{kl}
  dV}=\nonumber \\
  =\frac{1}{2}\left(\int\limits_V {C_{ijkl} A_{ijmn} A_{klrq} dV}
   \right) \Omega _{mn} \Omega _{rq} ,
\end{eqnarray}
  where   $C_{ijkl}$ is a tensor of elastic modules. When writing down kinetic
energy as
\begin{equation}\label{3}
  T = \frac{1}{2}\int\limits_V \rho  \dot u_i^2 dV =
  \frac{1}{2}\int\limits_V {\rho B_{ijk} \dot \Omega _{jk} B_{iml} \dot \Omega _{ml} dV}
\end{equation}
  we obtain the following expression for a Lagrangian
\begin{equation}\label{4}
  L = T - F = \frac{1}{2}\rho \tilde B_{jkml} \dot \Omega _{jk} \dot
  \Omega _{ml}  - \frac{1}{2}\tilde C_{sqrp} \Omega _{sq} \Omega _{rp} ,
\end{equation}
where
\begin{eqnarray}
 \tilde B_{jkml}\equiv \int\limits_V {B_{ijk}B_{iml}
  dV}; \nonumber \\  \\
 \tilde C_{mnrn}  \equiv \int\limits_V {C_{ijkl} A_{ijnm}
  A_{klrn} dV}.\nonumber
 \end{eqnarray}
The Lagrangian (4) leads to the equation of motion for $\Omega_{ij}$
\begin{equation}\label{5}
  \rho \tilde B_{ijml} \ddot \Omega _{ml}  + \tilde C_{ijrp}
   \Omega _{rp}  = 0.
\end{equation}

Now using the equations of motion (5) we may obtain the
characteristic combined equations for definition of natural
frequencies of oscillations of a sample
\begin{equation}\label{6}
\left( { - \omega ^2 \rho \tilde B_{ijml}  + \tilde C_{ijml} }
\right)\Omega _{ml}^0  = 0.
  \end{equation}

 For the analysis of experimental results  \cite{23} we shall choose
 a sample in the shape of the elliptic cylinder. It will allow to take
 into account  all characteristic dimensions of  experimentally studied samples
 \cite{23} and to derive  an expression for $\Omega_{ij}$  in the closed
 form, that essentially  simplifies  evaluation of  resonant frequencies.
Generally the stress tensor is determined by the following
equations and boundary conditions \cite{36}:
\[\addtocounter{equation}{1}
\raggedright\hspace{30pt}\left\{
  \begin{array}{llrrrrr}
   \frac{\partial \sigma _{ij}}{\partial x_k }=0,& (a)\\
 \delta _{ikn} \delta _{ilm} S_{mnpq} \frac{\partial ^2
 \sigma _{pq} }{\partial x_k \partial x_l }=0,&  (b)& &&&
 \hspace{6pt}(\arabic{equation}) \\
 \sigma _{ij} n_j  = p_i, & (c)
 \end{array}
\right.%
\]
  where $n_j$   is the normal unit vector to a surface of a sample, $p_j$  is
  the force acting on unit area of a sample. For the case of an
isotropic elliptic cylinder  the $\sigma_{ij}$ stress tensor   is a linear function of
coordinates \cite{35} and the equations (7b) are satisfied identically. Thus, the
stress tensor is determined only by the equations (7a), which, generally speaking,
have the same form as for isotropic, and an anisotropic material. Hence, the tensor of
strains and for the anisotropic elliptic cylinder will be a linear function of
coordinates and may be found on the basis of the analysis of boundary conditions and
has the following form
\begin{equation}
 u_{ij}  = \frac{{2\left| {\vec M}
\right|}}{{\pi a^2 b^2 }}\left[ {\left( {\frac{b}{a}} \right)x_1
S_{ij23}  - \left( {\frac{a}{b}} \right)x_2 S_{ij13} } \right],
\end{equation}
where $\vec{M}$  is  the torsional moment; $S_{ijkl}$  is  the tensor of elastic
compliances in laboratory system of coordinates and is related to the tensor in
crystallographic system through the transformation matrix  as $S_{i'j'k'l'}=A^{i}_{i'}
A^{j}_{j'}A^{k}_{k'}A^{l}_{l'}S_{ijkl}$
     By using (1) the  tensor $\Omega_{ij}$  may be written as
    \[\addtocounter{equation}{1}
    \begin{array}{l }
   \Omega _{ij}  = \frac{2\left|{\vec M}\right|}{\left(\pi a^2
b^2\right)}\times
 \nonumber \\ \\
 \left\| {\begin{array}{*{20}c}
   { - \left( {\frac{a}{b}} \right)S_{1313} } &
   { - \left( {\frac{a}{b}} \right)S_{2313} } &
   { - \left( {\frac{a}{b}} \right)S_{3313} }  \\
   { - \left( {\frac{b}{a}} \right)S_{1323} } &
   { - \left( {\frac{b}{a}} \right)S_{2323} } &
    { - \left( {\frac{b}{a}} \right)S_{3323} }  \\
   \begin{array}{r}
 \left( {\frac{b}{a}} \right)S_{1223}  +  \\
  + \left( {\frac{b}{a}} \right)S_{1113}  \\
 \end{array} & \begin{array}{l}
 \left( {\frac{b}{a}} \right)S_{2223}  +  \\
  + \left( {\frac{b}{a}} \right)S_{1213}  \\
 \end{array} & \begin{array}{l}
 \left( {\frac{b}{a}} \right)S_{2323}  +  \\
  + \left( {\frac{b}{a}} \right)S_{1313}  \\
 \end{array}  \\
\end{array}}
 \right\|\\
 \hspace{217pt}(\arabic{equation})
\end{array}\]

   In the case of $\vec u \bot \vec c,\vec k||\vec c$ the
equation of motion for $\Omega_{ij}$  (6) leads to the following
expression for natural frequencies of sample oscillations
\begin{equation}
  \omega^2=\frac{1}{\rho L^2}\frac{C_{1313} C_{2323}}{\left[
  \left(\frac{b}{a}\right)C_{1313}+\left(\frac{a}{b}\right)
  C_{2323}\right]}\frac{12ab}{a^2+b^2},
      \end{equation}
where  $\rho$ is  the density of a sample,  L is  the length of a sample along a
torsion axis;   $C_{ijkl}$ is the tensor of elastic modules expressed in laboratory
system of coordinates. In the case  $\vec k \bot \vec c,\vec u||\vec c$ the
characteristic equation for resonant frequencies of a sample have been obtained in a
similar way and because of complexity has been solved by numerical methods.
\mysubsection{Orientations $\vec u||\vec c, \vec k||\vec c; \,\vec u||\vec c,\vec k
\bot \vec c\,$}

For the given orientation of laboratory system of coordinates longitudinal vibrations
are raised in a sample. Calculation of natural frequencies of longitudinal vibrations
of a sample was performed on the basis of the equation of motion for a vector of
strains:
\begin{eqnarray}
   - \frac{{\partial ^2 u_3 }}{{\partial t^2 }} + \frac{{C_{ijkl}
  \Pi _{ij} \Pi _{kl} }}{{\rho \left( {1 + \Pi _{13}^2  + \Pi _{23}^2 }
   \right)}}\left( {\frac{{\partial ^2 u_3 }}{{\partial x_3^2 }}}
   \right)+ \nonumber \\
   + \frac{{S_\Delta ^2 }}{{12}}\frac{{\frac{{\vartheta _{11}^2  +
   \vartheta _{12}^2 }}{{b^2 }} + \frac{{\vartheta _{12}^2  + \vartheta
   _{22}^2 }}{{a^2 }}}}{{1 + \Pi _{13}^2  + \Pi _{23}^2 }}\frac{{\partial
   ^4 u_{_3 } }}{{\partial t^2 \partial x_3^2 }} = 0;
   \end{eqnarray}
where the $\Pi_{ij}$ tensor   is determined on the basis of the condition $u_{ij}  = -
\Pi _{ij} u_{33} $ ; $\vartheta _{11} \equiv \Pi _{13}^2  - \Pi _{11} ;\vartheta _{12}
= \vartheta _{21}  \equiv \Pi _{13} \Pi _{23}  - \Pi _{12} ;\vartheta _{22} \equiv \Pi
_{23}^2  - \Pi _{22} ;S_\Delta  $ is the cross-sectional area of a sample. The
equation (11) is obtained by a variation method and for  isotropic materials coincides
with the Relay equation  \cite{38}.
     The harmonic solution of the equation (11) can be deduced by the
      method of partitioning variables and  results in   the following
      expression for  natural frequencies of oscillations of a sample
\begin{eqnarray}
  \omega _n  = \left( {\frac{{2\pi n}}{L}} \right)%
  \sqrt {\frac{{C_{ijkl}\Pi _{ij} \Pi _{kl} }}{\rho%
  }}%
  \times\,\,\,\,\,\,\,\,\,\,\,\,\,\,\,\,\,\,\,\nonumber\\ \frac{1}{{\sqrt {1 + \Pi _{13}^2  +%
   \Pi _{23}^2  + \left( {\frac{{2\pi n}}{L}} \right)^2 \frac{{S_\Delta  }}%
   {{12}}\left( {\frac{{\vartheta _{11}^2  + \vartheta _{12}^2 }}{{b^2 }}%
    + \frac{{\vartheta _{12}^2  + \vartheta _{22}^2 }}{{a^2 }}} \right)} }},\nonumber
    \hspace{-20pt}\\%
    \end{eqnarray}
where n is an integer.

Let's analyze  temperature dependencies of the natural frequencies $\omega(T)$,
calculated on the basis of the expressions obtained above .
 First of all, one additional remark need to be made.
Recently we have  shown \cite{30}, that the $C_{3333} (T)$ temperature dependencies
observed  in \cite{22} may be qualitatively described within the framework of model of
a strongly correlated bistable sublattice and, hence, simulation of  natural
frequencies $\omega(T)$  of a sample also may be performed within the framework of the
given model. However here I shall act as follows. At calculation  of temperature
dependencies of natural frequencies it is convenient to choose as a modelling
temperature dependence of $C_{3333}(T)$ the $C_{3333}(T)$ dependence  measured in
\cite{22} , that enable not only to simulate temperature dependencies but in details
allows to compare results of works \cite{22} and \cite{23}. The calculated temperature
dependencies $\omega$(T) are depicted in figure 1. For all investigated orientations
calculated and experimental dependencies $\omega$(T)  are in excellent agreement.
 I pay attention, that for the orientation $\vec{u}\|\vec{c},\vec{k}\perp\vec{c}$
 a strong temperature dependence of $\omega$(T) is caused by  excitation of bending
 oscillations of  a sample  which natural frequency  is determined by  the $C_{3333}(T) $ module .
 \mysection{ Analysis of abnormal temperature dependencies of ultrasonic
velocity  in polycrystal samples}

For calculation of  effective values of a tensor of elastic
\end{multicols}
\begin{figure}[hbt]
      \includegraphics[width=15cm,height=15cm]{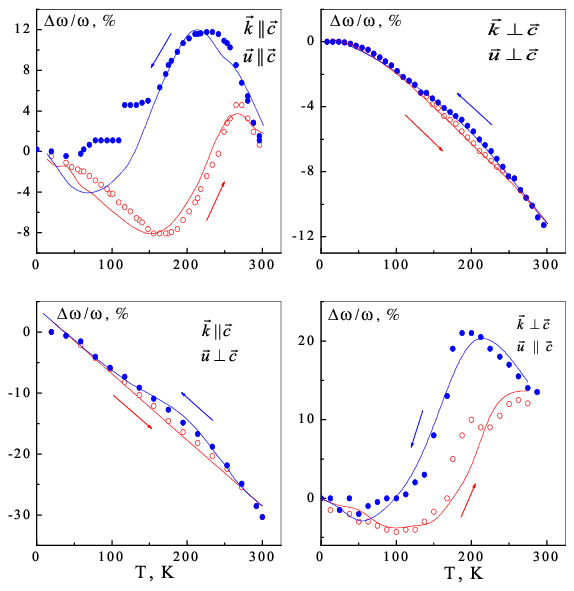}
      \centering
              \end{figure}
\begin{changemargin}{1cm}{1cm}
 {\small \vspace{-42pt}\noindent Figure 1. Temperature dependencies of the relative resonant
frequency $(\omega(T)- \omega(T_0))/\omega(T_0)$ of oscillations of a single crystal
    sample for various orientations of the wave vector $\vec{k}$ and the vector of
    displacement $\vec{u}$ of a ultrasonic wave . Lines represent simulations;
    points are experimental results \cite{23}; the direction of change of temperature is
    shown by arrows}
\end{changemargin}
 \begin{multicols}{2}\noindent
modules of polycrystal materials,  it is necessary to know
  topological type of the heterogeneous medium. From topological point of view all
  heterogeneous mediums may be subdivided on matrix
and statistical mediums. In matrix mediums the main  phase envelops all other phases
and forms connected space at any volume concentration of the main phase. In
statistical medium all phases are distributed by a casual fashion not forming regular
structures.
 For definition of topological type of the polycrystal medium
we shall break  volume of a sample on  elementary cubic regions (domains) with the
characteristic size $L_{0}$ ,  equal to the average size of a crystal grain. Let the
crystal grain has orthorhombic symmetry. Each separate crystal grain is oriented
arbitrarily relative to the chosen axis $x_{3}$ determining  propagation of an
ultrasonic wave. All set of domains of dissection breaks up to   three subsets in
which domains with orientation of   crystallographic  axes of a crystal grain
($\vec{a}, \vec{b}, \vec{c}$) belonging to the intervals:
\begin{eqnarray} Intr(\vec a) \in \left\{ {0 \le \angle (\vec
a\,\vec x_3 ) \le \frac{\pi }{4};\frac{{7\pi }}{4} \le \angle
(\vec a\,\vec x_3 ) \le 2\pi } \right\}  ;\nonumber \\Intr(\vec c)
\in \left\{ {0 \le \angle (\vec c\,\vec x_3 ) \le \frac{\pi
}{4};\frac{{7\pi }}{4} \le \angle (\vec c\,\vec x_3 ) \le 2\pi }
\right\}  ;\nonumber \\ Intr(\vec b) \in \left\{ {0 \le \angle
(\vec b\,\vec x_3 ) \le \frac{\pi }{4};\frac{{7\pi }}{4} \le
\angle (\vec b\,\vec x_3 ) \le 2\pi } \right\}.
\end{eqnarray}

The given splitting  can be performed always as it is possible to show that for Euler
angles the given relation is always fulfilled:  \begin{equation} \left( {\frac{\pi
}{4} \le \theta \le 0} \right) \Rightarrow
   \left( {\varphi  \ge \frac{\pi }{4},\psi  \ge
\frac{\pi }{4}} \right). \end{equation}

It is obvious, that for the chosen subset of domains the distribution function of
orientations of crystal grains $f(\theta)$  has a maximum in the intervals (13). And,
texture of a sample is determined by a relation of numbers of filling of the subsets
defined above. Further, for the given subset of domains we  replace the tensor of
elastic modules of each domain with average value of the tensor of elastic modules
$<C_{ijkl}>_\eta( \eta = Intr(\vec a),\,Intr(\vec b),Intr(\vec c) )$
 \cite{39}
  \begin{eqnarray}
\left\langle {C_{ijkl} } \right\rangle  = \frac{1}{{8\pi ^2 }}\times\nonumber \hspace{50pt}\\
\times\int\limits_0^{2\pi } {\int\limits_0^{2\pi } {\int\limits_0^\pi {C_{ijkl} \left(
{\theta ,\varphi ,\psi } \right)f} } } \left( {\theta ,\varphi ,\psi } \right)\sin
(\theta )d\varphi d\psi d\theta . \end{eqnarray} When performing  concrete
calculations the distribution function $f(\theta,\psi,\varphi)$ was chosen  in the
form $f(\theta,\psi,\varphi)=[\cos(\theta)]^{2n}$ [39], n - parameter (at
$n\rightarrow0$ or $n\rightarrow\infty$ we have random or textured distribution of
orientations of crystal grains, respectively). As a result, polycrystal material can
be considered as the three-phase heterogeneous medium. Topological type of medium is
determined by a relation of the relative volumes $V_{\vec{a}}$ , $V_{\vec{b}}$
,$V_{\vec{c}}$ and the percolation threshold $X_{c}$( $X_{c}=0.307 \div 0.325$ for the
site problem on a cubic lattice \cite{40}). So the conditions for matrix and
statistical mediums have the form $V_{\vec{a}}<X_{c}$(matrix medium concerning phases
$V_{\vec{b}}$ ,$V_{\vec{c}}$) and $V_{\vec{a}}$ , $V_{\vec{b}}$ ,$V_{\vec{c}}> X_{c}$,
respectively.
 \mysubsection{ Finely grained dense polycrystal ceramics}

For $\rm YBa_{2}Cu_{3}O_{7-\delta}$ compound  at $\delta\approx0$ crystal grains has
orthorhombic symmetry. Orientation of crystal grains is arbitrary for a nontextured
sample and for relative volumes of separate phases  we have the following condition
$V_{\vec{a}}\approx{V_{\vec{b}}}\approx{V_{\vec{c}}}\approx\frac{1}{3}$. The
calculation of average values of a tensor of elastic modules has shown, that for
compound $\rm YBa_{2}Cu_{3}O_{7-\delta}$ we have the following relation
$<C_{1111}>\approx<C_{2222}> $. Hence, in this case polycrystal material is matrix
medium and  the matrix is formed by crystal grains with relative volume
${V_{\vec{a}}+V_{\vec{b}}>X_c}$. For the matrix material the effective elastic modulus
may be calculated using the virial approach based on decomposition of an effective
tensor of elastic modules in a series on relative volume of a phase ${V_0<X_c}$ and in
linear approach expression for the effective module looks like \cite{39}
  \begin{equation}
C^*  = \left\langle {C_{1111,2222} } \right\rangle \left[ {1 +
\frac{{3\left( {\left\langle {C_{3333} } \right\rangle  -
\left\langle {C_{1111,2222} } \right\rangle }
\right)}}{{2\left\langle {C_{3333} } \right\rangle  + \left\langle
{C_{1111,2222} } \right\rangle }}V_{\vec c} } \right]
  \end{equation}
Using expression (14) and average values of a tensor of elastic modules (15) we have
calculated temperature dependence of ultrasonic velocity for dense finely grained
polycrystal materials. At calculation was considered, that  the temperature
dependencies of elastic modules of separate crystal grains and single crystals are
identical. Results of calculation are depicted in figure 2. The main feature of
temperature dependencies in this case is absence of abnormal behavior of temperature
dependencies of ultrasonic velocity though the elastic modules  $C_{3333}$ of separate
crystal grains had pronounced hysteretic temperature dependence. Let's pay attention,
that the calculated temperature dependencies are in a good agreements with
experimental ones \cite{20}.

 Thus, for not textured finely grained   polycrystal
materials   the temperature dependencies of velocity of ultrasound are determined  by
average values of the $<C_{1111}>, <C_{2222}>$ crystalline elastic modules.
 \mysubsection{Textured polycrystal materials}

In the case of textured polycrystal material for the relative volume occupied by the
crystal grains  oriented, for example, in $\vec{c}$ direction the following condition
is fulfilled $V_{\vec{c}}>V_{\vec{a}}, V_{\vec{b}}$ and the relative volumes
$V_{\vec{a}}, V_{\vec{b}}$ are much less than the percolation threshold $X_c$. In this
case we have the matrix topology of polycrystal medium, but the matrix is formed by
crystal grains with average value of the elastic modulus $C_{3333}$ . For calculation
of temperature dependencies expression (14) can be used in which the following
replacement $<C_{3333}>\Leftrightarrow<C_{1111}, C_{2222}>$ have been performed.
Calculated temperature dependencies of ultrasonic velocity for textured polycrystal
samples with relative volume $V_c\approx0.9$ are depicted in figure 2b. In this case
hysteretic behavior of \vT is observed. Similar temperature dependencies were observed
experimentally for textured samples in \cite{18}.
 \mysubsection{ Materials with topology of a statistical mixture}

 Let's consider  polycrystal materials, when for relative volumes
of separate sets of crystal grains the following condition  is satisfied
$V_{\vec{c}}\approx0.5>X_c; V_{\vec{a}}\cup{V_{\vec{b}}}\approx0.5>X_c$. In general
case for calculation of an effective tensor of elastic moduli detailed distribution of
inhomogeneities in a sample must be known \cite{35}. However temperature dependence of
ultrasonic wave velocity can be analyzed on the basis of the percolation theory.
Really, from the percolation theory  follows immediately \cite{40} that the whole
volume of a sample of a statistical mixture can be divided into two connected clusters
with  different orientation of crystal grains $Intr(\vec{c})$ and $Intr(\vec{a})$,
$Intr(\vec{b})$, respectively. And, apparently, that properties of a statistical
mixture should be symmetrical concerning replacement
$<C_{3333}>\Leftrightarrow<C_{1111}, C_{2222}>$. For a wave length of ultrasound the
following relation $\lambda_{sound}<L_{cluster}\approx{L_{sample}}$ is fulfilled and
elastic waves propagate in both clusters. When an elastic plane wave propagates in
nonuniform material wave front will be deformed and  the spatial modulation of
amplitude will appear. It will take place even in a case when end faces of a sample
are absolutely parallel and flat. The value of
\end{multicols}

 \begin{figure}[hbt]
      \includegraphics[width=9cm,height=12cm]{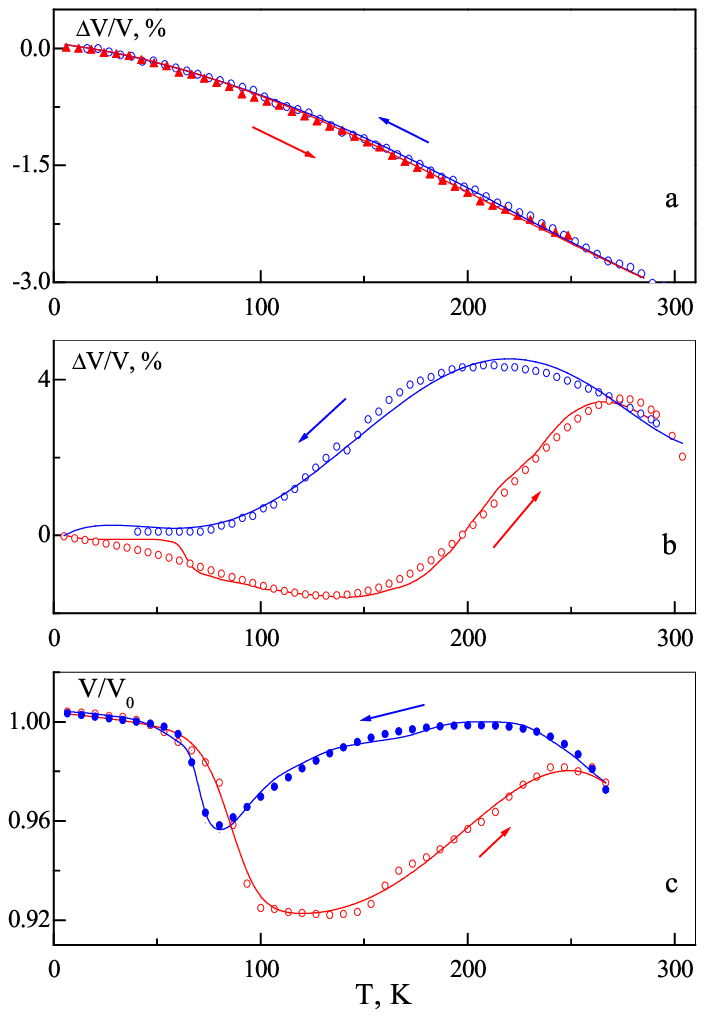}
      \centering
   \end{figure}
  \begin{changemargin}{1cm}{1cm}
 {\small \vspace{-32pt}\noindent Figure 2. Temperature dependencies of the relative velocity
 of ultrasound  ($(v(T)-v(T_0))/v(T_0)$ for a,b and $V(T)/v(T_0)$ for c)
in $YBa_2Cu_3O_{7-\delta}$ polycrystal samples.  Lines represent simulations; points
are experimental results: a - for dense, small crystal grains sample \cite{20};b -
textured sample \cite{18}; c - sample with relatively large crystal grains
(${\sim}50{\mu}m$) \cite{19}; the direction of change of temperature is shown by
arrows}
\end{changemargin}
\begin{multicols}{2}
 \noindent  a signal received by a detector sensitive to amplitude
of strains  will be proportional to a wave amplitude averaged over the detector
surface
\begin{eqnarray}
u_0  = k_0 \left\langle A \right\rangle _S ;\left\langle A
\right\rangle _S  = \frac{1}{S}\int\limits_S {\sum\limits_i {} A_i
(x,y)dxdy} ,
\end{eqnarray}
 where summation is carried out over all connected clusters.  Elastic waves
propagating in connected clusters are detected as a single wave with effective
half-width and effective standing of a maximum $<t>$ . The average value $<t>$ is
determined by a relationship of amplitudes $A_i$ and times $t_i$ of elastic waves
propagating in each cluster and in a linear approach may be represented as the
ensemble average
\begin{equation}
\left\langle t \right\rangle  = \sum\limits_i {\left( {\frac{{A_i
}}{{\left\langle A \right\rangle _s }}} \right)^\gamma  t_i } ,
\end{equation}
 where $\gamma$ is the parameter of the given detector system. Expressing the amplitude of a
 passed wave as $A_i  = A_0 \exp \left( { - \alpha _i L} \right)$ the following
expression for average velocity in a case of a statistical mixture
can be obtained
 \begin{equation}
\left\langle {\mathop{v(T)}\nolimits}  \right\rangle  = \frac{L}{{\left\langle t
\right\rangle }} = \frac{{\left( {\frac{{S_0 }}{{S_{} }}} \right)^\gamma  \left[ {1 +
\exp \left( {\alpha _1 L_1  - \alpha _2 L_2 } \right)} \right]^\gamma  }}{{1 + \left(
{\frac{{{\mathop{v}\nolimits} _{\rm 1}(T) }}{{{\mathop{v}\nolimits} _{\rm 2}(T) }}}
\right)\exp \left[ {\gamma \left( {\alpha _1 L_1  - \alpha _2 L_2 } \right)}
\right]}}{\mathop{ v}\nolimits} _{\rm 1}(T),
 \end{equation}
where  $v_i, \alpha_i$ are the velocity and the decay factor of an ultrasound wave,
$L_i$ is the average size of  a cluster; $S_0$  is the average sectional area of a
cluster; $S$ is the effective area of a detector. As follows from equation (19) ,
$<v(T)>$ temperature dependence is being determined  by $v_1(T)$ or $v_2(T)$
temperature dependence depending on a relation of $\alpha_i(T)$ decay factors. I have
simulated $<v(T)>$ temperature dependencies of ultrasonic wave velocity taking into
account experimentally measured temperature dependencies  of a decay factor \cite{19,
22} (see Figure 2). Temperature dependence at high temperatures is determined by
temperature dependence of ultrasonic wave velocity in the cluster with orientation of
crystal grains $Intr(\vec{c})$. At decreasing of temperature the decay factor of
ultrasound  in the cluster with orientation of crystal grains $Intr(\vec{c})$ is
sharply increasing while in the
\end{multicols}
\begin{figure}[hbt]
\includegraphics[width=12cm,height=10cm]{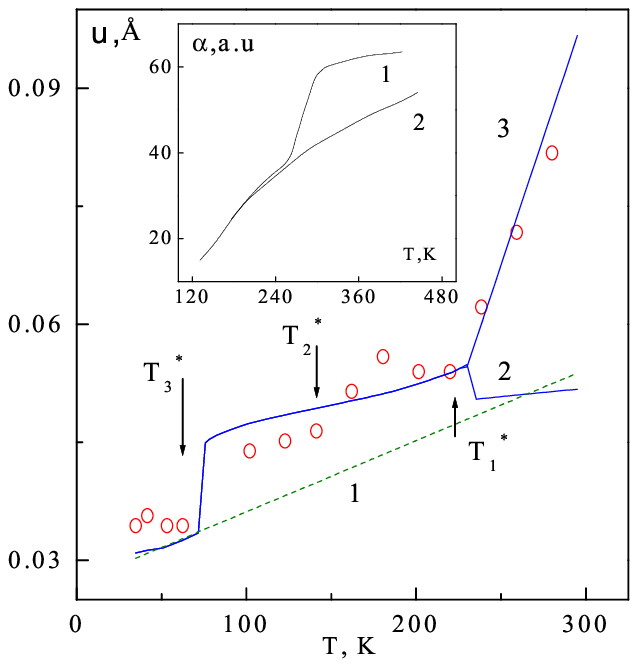}
\centering
\end{figure}
\begin{changemargin}{1cm}{1cm}
{\small \vspace{-35pt} \noindent Figure 3.Temperature dependence of incoherent lattice
fluctuation in $\rm YBa_2Cu_3O_{7-\delta}$ crystals.
 Points are experimental results \cite{45}. Lines are calculated temperature
 dependencies: 1 -  Debye model; 2,3 - model of strongly correlated sublattice without and with
 taking into account disordering of oxygen atoms in Cu(1)-O(1) chains. On the inset - expansivity versus
 temperature of  untwined $\rm YBa_2Cu_3O_{7-\delta}$ single crystals for $\delta \sim 6.95$ (curve 1)
  and $\delta \sim 7$(curve 2) \cite{28}.}
\end{changemargin}
\begin{multicols}{2}
\noindent cluster with orientation $Intr(\vec{a})$, $Intr(\vec{b})$ is decreasing
\cite{19, 22} and as a result velocity of ultrasound will be determined by the cluster
of the type $Intr(\vec{a})$, $Intr(\vec{b})$. The given temperature dependence of \vT
represents in some sense a combination of the temperature dependencies considered
above for textured and finely grained materials. It is interesting to note, that a
similar sort of temperature dependencies were observed experimentally in samples with
large crystal grains (50 micron) \cite{19} (see fig. 2). Thus, quantitative analysis
of experimentally observed hysteretic temperature dependencies of ultrasonic wave
velocity in $\rm YBa_2Cu_3O_{7-\delta}$ polycrystal materials
 (finely grained, textured, macrocrystalline) at frequencies
from 100 kHz up to 20 MGz has shown, that as well as in  case of single crystal
samples, the hysteretic behavior is determined by  hysteretic temperature behavior of
the modulus $C_{3333}$ while the other elements of a tensor of elastic moduli have no
abnormal temperature dependence.
 \mysubsection{On microscopic mechanism of hysteretic behavior}
 The fact of existence of  strong anisotropy of  hysteretic behavior of elastic modulus
 is important from the point of view of clearing up of the microscopic mechanism of so
 unusual behavior.
 Really,  the essential contribution to anisotropy of a tensor of  elastic modulus is given
  by interactions of atoms of the nearest coordination spheres of crystal.
Having expressed a tensor of elastic moduli as a function of constants of pair
interactions of atoms  in $\rm YBa_2Cu_3O_{7-\delta}$ crystal  it can easily be shown
that the abnormal temperature behavior of the elastic modulus $C_{3333}$   is caused
by change with temperature of  force constants of the apical oxygen atoms. In used
approach the modulus $C_{3333}$ is determined by the interaction of apex oxygen atoms
with copper and barium atoms but as abnormal temperature behavior of the elastic
moduli $C_{1111}$, $C_{2222}$  is not observed, a sole opportunity is temperature
renormalization of the O(4) oxygen - copper force constants. The given conclusion is
consistent with experimental results \cite{41} (also references in it) \cite{42,43,44}
in which features in dynamics of O (4), O(1) oxygen atoms in $\rm
YBa_2Cu_3O_{7-\delta}$ were observed also. From the point of view of study of dynamics
of atoms  in $\rm YBa_2Cu_3O_{7-\delta}$ the
 recently published article \cite{45} is of interest.
In \cite{45} it is shown that in the temperature region 100 - 200 K incoherent lattice
fluctuations  essentially surpass thermal ones. Growth of incoherent lattice
fluctuations $u(T)$ were observed at temperature $T^*_1\approx200 K$ and  the critical
temperature $T^*_1$ is attributed in \cite{45} to the formation of charge ordering. It
is very interesting that the temperature interval in which incoherent lattice
fluctuations exceeds thermal coincides with the interval in which hysteretic behavior
of speed of ultrasound is observed. And the temperature   of formation of charge
ordering $T^*_1$ corresponds to the temperature of opening of a hysteresis loop. In
this connection I shall pay attention to the experimental result \cite{46} where
hysteretic behavior of ultrasonic wave velocity  in CuO crystals  was observed. And
recently in this compound charge-ordered domains were revealed \cite{47}.
Charge-ordered domains with necessity results in occurrence of correlated states of
lattice degrees of freedom, that may be the reason of formation in a crystal of
strongly correlated  groups of atoms. Recently, we offered  the model of strongly
correlated sublattice for an explanation of hysteretic behavior of ultrasonic wave
velocity and thermal conductivity in $\rm YBa_2Cu_3O_{7-\delta}$ crystals  \cite{30}.
Within the framework of the given model I have calculated temperature dependence
$u(T)=\sqrt{\sigma(T)}$ (for more details see \cite{30}). The calculated curve $u(T)$
describes experimental results for temperatures $T<T^*_1$. However at higher
temperatures on the experimental dependence $u(T)$ essential growth of incoherent
lattice fluctuations  is observed. The given growth of fluctuations  may be caused by
a disordering of oxygen atoms in the Cu(1)-O(1)  chains. Really, a quiet large anomaly
of a linear expansion coefficient was observed  in nearly optimally doped
($\delta$=0.68) of $\rm YBa_2Cu_3O_{7-\delta}$ untwined single crystals in the same
temperature region \cite{28}. But in fully oxygenated ($\delta$=7.0) crystals such
anomaly were absent (see inset in fig. 3). I have revaluated the $u(T)$ temperature
dependence with taking into account  a disordering of vacancies of atoms of oxygen
(Fig. 3, curve 3). One can see that the calculated temperature dependence well enough
describes experimental results.

Thus, the  analysis performed shows that observed experimentally incoherent lattice
fluctuations in $\rm YBa_2Cu_3O_{7-\delta}$ crystals may be caused by particular
dynamics of oxygen atoms and I can speculate that the hysteretic behavior of the
$C_{3333}$ elastic modulus  may be attributed to the formation of charge ordering at
temperatures $T<T^*_1$.
   \mysection{Conclusion}

The joint analysis of temperature dependencies of sample resonant frequencies and
ultrasonic wave velocity as in $\rm YBa_2Cu_3O_{7-\delta}$ single crystal samples with
various twinning structures and in polycrystal samples measured over a brode range of
frequencies (100 kHz $\div$ 20 MHz), shows strong anisotropy of hysteretic behavior of
elastic moduli. Namely, the $C_{3333}$  and in much smaller degree $C_{2323}$ moduli
have  hysteretic temperature dependence. While the moduli $C_{1111}, C_{2222},
C_{1212}$ and $C_{1313}$ do not show abnormal temperature dependence. The analysis of
a tensor of elastic moduli of $\rm YBa_2Cu_3O_{7-\delta}$   crystal on the basis of
the microscopic model which is taking into account interaction of atoms of the nearest
coordination spheres has shown that the anomalies in behavior of an elastic modulus
are caused by a temperature-dependent renormalization of force constants of apex
oxygen atoms  with Cu(1), Cu(2) copper atoms  and apparently are attributed to  the
formation of charge ordering in $\rm YBa_2Cu_3O_{7-\delta}$ crystals.

 \end{multicols}
\end{document}